\documentclass[12pt,psfig]{article}
\usepackage{amssymb}
\usepackage{amsmath,amsthm}
\usepackage{amsfonts}
\usepackage{graphicx}
\usepackage{braket}
\usepackage{graphics}
\usepackage{indentfirst}
\numberwithin{equation}{section}

\usepackage{hyperref}

\begin{document}
\begin{center}\Large\textbf{Gravitational and Dilatonic
Radiations from the Parallel Dressed-Dynamical
Unstable D$p$-Branes}
\end{center}
\vspace{0.75cm}
\begin{center}
{\large Hamidreza Daniali and \large Davoud Kamani}
\end{center}
\begin{center}
\textsl{\small{Department of Physics, Amirkabir University of
Technology (Tehran Polytechnic) \\
P.O.Box: 15875-4413, Tehran, Iran \\
e-mails: hrdl@aut.ac.ir , kamani@aut.ac.ir \\}}
\end{center}
\vspace{0.5cm}

\begin{abstract}

In the context of the bosonic string theory,
we shall extract the general radiation amplitude
of a massless closed string
from the interaction of two parallel unstable
D$p$-branes. The branes are non-stationary and
have been dressed by background fields.
The foregoing amplitude
will be rewritten for the massless state radiation from
the branes with the large distance.
Besides, we shall study the gravitational
and dilatonic radiations from this configuration.

\end{abstract}



\newpage
\section{Introduction}

Some significant physical results in string theory have been
obtained by investigating the D-branes and
their interactions \cite{1, 2}.
The boundary state formalism \cite{3}-\cite{16} is an
adequate technique for calculating the interaction of
the branes in the closed string channel.
The D-branes in the presence of the nonzero
background fields and dynamics
possess some appealing properties.
Therefore, by adding various background fields and dynamics
a general boundary state, corresponding to
a D$p$-brane, can be obtained.
The interaction of such branes has been widely studied
via the boundary state formalism \cite{17}-\cite{28}.

The D$p$-branes can radiate closed strings in a wide range of
configurations.
One of the most significant setups is the production of
closed strings from a single
unstable D$p$-brane \cite{29, 30, 31}.
The closed string radiation from a single unstable brane in
the presence of the various background fields has
been also studied \cite{32, 33}.
Besides, the supersymmetric form of the closed string
radiation was investigated \cite{34}.
The closed string radiation from the interacting
D-branes is another setup for the production of
particles in the closed string spectrum \cite{35, 36}.
We shall apply the latter case for a generalized
configuration of the interacting branes.

The background fields and dynamics of the branes
motivated us to study the effects
of these variables on the radiation of a
massless closed string
from the interacting branes.
Thus, in this paper, in the context
of the bosonic string theory,
we study a massless closed string
radiation via the interaction of two parallel
unstable D$p$-branes. The branes have
tangential dynamics and they have been dressed by
the quadratic tachyonic fields,
the Kalb-Ramond field
and the $U(1)$ gauge fields in a specific gauge.
The boundary state formalism will be used to compute the
radiation amplitude.
Hence, by entering an appropriate vertex operator in
the worldsheet of the mediated closed string between
the branes, we obtain a radiated closed string.
We shall accurately utilize the eikonal approximation,
which ignores the recoil of the branes.
At first, we compute the amplitude for
radiating a general massless closed string.
Afterward, we focus on the graviton and dilaton
emission from large distance branes.
We shall see that the resultant particle is radiated
from one of the branes (a Bremsstrahlung-like process)
or from the middle area between the branes.
Note that the interaction between two
branes, with a large distance, entirely occurs
through the exchange of the massless closed strings.

This paper is organized as follows. In
Sec. \ref{200}, we shall introduce the boundary state,
which is associated with a
dressed-dynamical unstable D$p$-brane.
In Sec. \ref{300}, the radiation amplitude of a generic
massless closed string,
created from the interaction of
two parallel D$p$-branes, will
be computed. This amplitude will be deformed to
describe radiation from the large
distance branes. In Sec. \ref{400},
the emission of the graviton and dilaton from
the large distance branes, with a specific
set of the background fields and
dynamics, will be calculated.
Sec. \ref{500} is devoted to the conclusions.

\section{Dressed-dynamical unstable
D$p$-brane: the boundary state}
\label{200}

In order to compute the boundary state, corresponding to a
dressed-dynamical unstable D$p$-brane,
one should start with the following string sigma-model
action
\begin{eqnarray}
\label{2.1}
S_\sigma =&-&\frac{1}{4\pi\alpha'} {\int}_\Sigma
{\rm d}^{2}\sigma \left(\sqrt{-g}g^{ab}
\mathcal{G}_{\mu\nu}\partial_a X^{\mu}\partial_b
X^{\nu}+\varepsilon^{ab} B_{\mu\nu}
\partial_a X^{\mu}\partial_b
X^{\nu}\right)
\nonumber\\
&+&\frac{1}{2\pi\alpha'} {\int}_{\partial\Sigma}
{\rm d}\sigma \left( A_\alpha
\partial_{\sigma}X^{\alpha}+
\omega_{\alpha \beta}\mathfrak{J}^{\alpha\beta }_{\tau }
+\mathcal{T}(X^\alpha)\right),
\end{eqnarray}
where $\Sigma$ is the worldsheet of the closed string
and $\partial\Sigma$ represents its boundary.
The spacetime indices,
i.e, $\mu,\nu \in \{0,1,\cdots , 25\} $ are split into the
indices for the worldvolume directions
$\alpha, \beta \in \{0 , 1,\cdots ,p \}$,
and the ones for the perpendicular directions
$i, j \in \{p+1 , \cdots ,25 \}$.
In addition, the worldsheet directions
are labeled by $a,b\in\{0,1\}$.
We employ a constant Kalb-Ramond field
$B_{\mu\nu}$, a $U(1)$ internal gauge field
$A_{\alpha}=-\frac{1}{2}F_{\alpha \beta }X^{\beta}$ with the
constant field strength $F_{\alpha \beta }$, and a
quadratic open-string tachyonic
field $\mathcal{T}=
\frac{1}{2}U_{\alpha\beta}X^{\alpha}X^{\beta}$,
where the matrix $U_{\alpha\beta}$ is symmetric
and constant. The antisymmetric
tensor $\omega_{\alpha \beta}$
$(\mathfrak{J}^{\alpha\beta}_{\tau})$
will be used for the angular velocity
(angular momentum density).
Hence, the explicit form of the dynamics term is given by
${\omega }_{\alpha \beta}\mathfrak{J}^{\alpha\beta}_{\tau}
= 2{\omega }_{\alpha \beta}X^{\alpha}{\partial }_{\tau
}X^{\beta}$.
We should note that in the presence of
the background fields on the worldvolume of the brane,
the Lorentz symmetry has been prominently
broken. Hence, the brane tangential dynamics
along its worldvolume directions is
meaningful.

By applying the flat spacetime and flat worldsheet,
and also by setting the variation of the action
to zero, the following boundary state
equations are conveniently extracted
\begin{eqnarray}
\label{2.2}
&~& \left[\left(\eta_{\alpha \beta }
+4{\omega }_{\alpha \beta}\right){\partial}_{\tau }X^{\beta}
+{\mathcal F}_{\alpha \beta}
{\partial }_{\sigma }X^{\beta}
+U_{\alpha \beta }X^{\beta }\right]_{\tau =0}\ \
 |\mathcal{B}\rangle\ =0,
\nonumber\\
&~& {\delta X}^i|_{\tau =0}|\mathcal{B}\rangle\ =0,
\end{eqnarray}
where we have defined the total field strength
as ${\cal{F}}_{\alpha \beta}=F_{\alpha \beta}-B_{\alpha \beta}$.

By utilizing the solution of the
equation of motion into the boundary state
equation, we can rewrite Eqs. \eqref{2.2}
in terms of the string oscillators. Besides,
the solution of the total boundary
state equations can be represented as follows
\begin{eqnarray}
{|\mathcal{B}\rangle}^{(\rm tot)}=\frac{T_p}{2}
{|\mathcal{B}\rangle}^{(\rm osc)}\otimes
{|\mathcal{B}\rangle}^{(0)}\otimes
{|\mathcal{B}\rangle}^{(\rm gh)},
\label{2.3}
\end{eqnarray}
where $T_p$ is the D$p$-brane tension, and
\begin{eqnarray}
\label{2.4}
{|\mathcal{B}\rangle}^{({\rm osc})}\ &=&\prod^{\infty }_{n=1}
{[\det Q_{(n)}]^{-1}}{\exp \left[-\sum^{\infty }_{m=1}
{\frac{1}{m}\left({\alpha }^{\mu }_{-m}\mathcal{S}_{(m)\mu \nu }
{\widetilde{\alpha }}^{\nu }_{-m}\right)}\right]\ }
{|0\rangle}_{\alpha }
\otimes{|0\rangle}_{\widetilde{\alpha }}\;,\\
{{\rm |}\mathcal{B}\rangle}^{\left(0\right)}&=&
\prod_i{\delta {\rm (}x^i}{\rm -}y^i{\rm )}
{\rm |}p^i{\rm =0}\rangle\;
\int^{\infty }_{{\rm -}\infty }
{\prod^p_{\alpha=0}\bigg{\{}{\rm d}p^{\alpha }}
\exp\Bigg[i{\alpha }^{{\rm '}}\Bigg(
{\left(U^{{\rm -}{\rm 1}}
{\mathbf A}\right)}_{\alpha \alpha }
{\left(p^{\alpha }\right)}^{{\rm 2}}
\nonumber\\
&+&\sum^{}_{\beta \ne \alpha}
{{\left(U^{{\rm -}{\rm  1}}
{\mathbf A}+{\mathbf A}^T U^{-1}\right)}_{\alpha \beta }
p^{\alpha }p^{\beta }}\Bigg)\Bigg]{\rm \ \ }
{\rm |}p^{\alpha }\rangle
\bigg{\}}, \ \ \ \ \label{2.5}\\
{|\mathcal{B}\rangle}^{(\rm gh)}
&=& \exp\left[\sum^{\infty }_{m=1}
{\left(c_{-m}{\tilde{b\ }}_{-m}-b_{-m}
{\tilde{c}}_{-m}\right)}\right]\frac{c_0+{\tilde{c}}_0}{2}
|q=1\rangle\ \otimes|\tilde{q}=1\rangle.
\label{2.6}
\end{eqnarray}
The coherent state method has been applied
to calculate the first equation.
The last state exhibits the conformal ghosts
contribution to the total boundary state.
In addition, we have defined the following matrices
\begin{eqnarray}
&~& Q_{(m){\alpha \beta }} =
{\mathbf A}_{\alpha \beta}-
{{\mathcal F}}_{{\mathbf \alpha }{\mathbf \beta
}}+\frac{i}{2m}U_{\alpha \beta },
\label{2.7}\\
&~& \mathcal{S}_{(m)\mu\nu}=(\Delta_{(m)\alpha \beta}\;
,\; -{\delta}_{ij}),
\label{2.8}\\
&~& \Delta_{(m)\alpha \beta} =
(Q_{(m)}^{-1}N_{(m)})_{\alpha \beta},\label{2.9}\\
&~& N_{(m){\alpha \beta }} = {\mathbf A}_{\alpha \beta}
+{{\mathcal F}}_{{\mathbf \alpha }{\mathbf \beta }}
-\frac{i}{2m}U_{\alpha \beta },\label{2.10}\\
&~&{\mathbf A}_{\alpha \beta}=\eta_{\alpha \beta}
+ 4\omega_{\alpha \beta}.
\label{2.11}
\end{eqnarray}
It should be mentioned that the normalization prefactor
in Eq. \eqref{2.4} comes from the disk partition function.

\section{General formulation for a massless state radiation}
\label{300}

In this section, we concentrate on the radiation of
a massless closed string
from the interaction of two dressed-dynamical
unstable D$p$-branes.
To achieve a generalized result,
distinct fields and different dynamics
are applied to each of
the branes. Thus, the fields and
dynamics of the first and second branes are labeled
by the subscripts (1) and (2), respectively.

Interaction between two D-branes,
in the closed string channel,
is elaborated by exchanging a
closed string between two boundary states.
The geometry of the worldsheet of the exchanged
closed string is a cylinder.
Let $\tau$ denote the coordinate along the length of the
cylinder, $ 0\le \tau \le t$, and $\sigma$ as the
periodic coordinate, i.e. $ 0 \le \sigma\le \pi$.
The radiation of a closed string is
described by inserting an appropriate vertex operator,
corresponding to the radiated string, into the interaction
amplitude.
From the mathematical point of view, one should calculate
the following amplitude
\begin{equation}
\label{3.1}
\mathcal{A} = \int_{0}^{\infty} dt \int_{0}^{t} d\tau \int_0^\pi
d \sigma \ ^{(\rm tot)}\langle \mathcal{B}_1| e^{-tH}
\mathcal{V}(\tau, \sigma) |\mathcal{B}_2\rangle^{(\rm tot)},
\end{equation}
where $H$ is the closed string Hamiltonian,
which comprises the matter and ghost parts.
Besides, $\mathcal{V}(\tau, \sigma)$
represents a vertex operator, associated with an
arbitrary massless string.
Let us exert the complex coordinate
$z=\sigma +i \tau$ and the derivative
$\partial = \partial_z$.
Thus, the vertex operator is given by
\begin{equation}
\label{3.2}
\mathcal{V}(z, \bar{z}) =
\epsilon_{\mu\nu} \partial X^\mu
\bar{\partial} X^\nu e^{ip \cdot X},
\end{equation}
where the momentum of the radiated closed string is
$p^\mu$ (with $p^\mu p_\mu =0$), and its
polarization tensor is $\epsilon_{\mu\nu}$.

By substituting the vertex operator \eqref{3.2}
into the amplitude \eqref{3.1}, we find
\begin{eqnarray}
\label{3.3}
\mathcal{A} &=& \dfrac{T_p^2}{4}
(2\pi)^{26} \epsilon_{\mu\nu}
\int_{0}^{\infty} dt \int_{0}^{t} d\tau
\int_0^\pi d \sigma \
\mathcal{D}(y_1, y_2) e^{4t}
\int_{-\infty}^{+ \infty}
\int_{-\infty}^{+ \infty}
\prod_{\alpha=0}^{p} dk^\alpha dk^{\prime \alpha}
\mathfrak{D} (k,k^\prime)
\nonumber \\
&\times& e^{-t \alpha^\prime k^2}
e^{- \tau \alpha^\prime (k^{\prime 2} - k^2)}
Z^{\rm (osc,g)} \langle
e^{ip\cdot X_{\rm osc}} \rangle
\Big\{\langle \partial X^\mu \bar{\partial} X^\nu
\rangle_{\rm osc} - \langle \partial
X^\mu p \cdot X\rangle_{\rm osc}
\langle \bar{\partial} X^\nu p\cdot X \rangle_{\rm osc}
\nonumber\\
&-& i \alpha^\prime k^\mu
\langle \bar{\partial} X^\nu p\cdot X \rangle_{\rm osc}
+ i \alpha^\prime k^\nu
\langle \partial X^\mu p\cdot X \rangle_{\rm osc} -
\alpha^{\prime 2}k^\mu k^\nu \Big\}, \ \ \
\end{eqnarray}
where the following quantities
originate from the zero-mode part of the amplitude
\begin{eqnarray}
\mathcal{D}(y_1, y_2) &\equiv&
\prod_{i=p+1}^{25} \delta(x^i - y_1^i)
\delta(x^i - y_2^i) \delta(p^i),
\label{3.4}\\
\mathfrak{D} (k,k') &\equiv&
\prod_{\alpha=0}^{p} \delta
\left( p^\alpha + k^{\prime\alpha} - k^\alpha\right)
\cr
&\times& \exp\Bigg[ -i \alpha^\prime
\Bigg( \sum_{\alpha = 0}^{p} \left[(U_{1}^{-1}
\mathbf{A}_1)_{\alpha\alpha} (k^\alpha)^2
-(U_{2}^{-1}\mathbf{A}_2)_{\alpha\alpha} (k'^\alpha)^2
\right]
\cr
&+& 2\sum_{\beta\ne\alpha}\left[(U_1^{-1} \mathbf{A}_1
)_{\alpha\beta}k^\alpha k^\beta
- (U_2^{-1} \mathbf{A}_2 )_{\alpha\beta}
k'^\alpha k'^\beta \right]
\Bigg) \Bigg] .
\quad\
\label{3.5}
\end{eqnarray}
Besides, the oscillating part of the
partition function $Z^{\rm (osc,g)}$
and the correlators
in Eq. \eqref{3.3} possess the following definitions
\begin{eqnarray}
\label{3.6}
Z^{\rm (osc,g)}= \ ^{\rm(osc)}\langle \mathcal{B}_1|
e^{-t H_{\rm(osc)} }
|\mathcal{B}_2\rangle^{\rm(osc)}\;\;
^{\rm{(g)}}\langle \mathcal{B}_1|
e^{-t H_{\rm(g)} }
|\mathcal{B}_2\rangle^{\rm{(g)}},
\end{eqnarray}
\begin{eqnarray}
\label{3.7}
\langle \mathcal{O}(\sigma , \tau)\rangle_{\rm osc}
\equiv \dfrac{^{\rm(osc)}\langle \mathcal{B}_1|
e^{-t H_{\rm (osc)} }\mathcal{O} (\sigma , \tau)
|\mathcal{B}_2\rangle^{\rm(osc)}}{^{\rm(osc)}
\langle \mathcal{B}_1| e^{-t H_{\rm (osc)} }
|\mathcal{B}_2\rangle^{\rm(osc)}}\;.
\end{eqnarray}
According to Eq. \eqref{3.3},
$\mathcal{O}(\sigma, \tau)$ takes four values.

The first and second integrals in
Eq. \eqref{3.3} guarantee that all points
of the worldsheet cylinder have been considered.
Let us introduce another proper time, i.e. $t'= t-\tau$.
It enables us to modify the integrations as
\begin{equation}
\int_{0}^{\infty}dt \int_{0}^{t} d\tau
= \int_{0}^{\infty} d\tau \int_{0}^{\infty}
dt^\prime.
\label{3.8}
\end{equation}
Note that $t^\prime$ ($\tau$) denotes the proper time
of the closed string, which is radiated from the
brane with the boundary state $|\mathcal{B}_1\rangle$
($|\mathcal{B}_2\rangle$). This means that $t^\prime =0$
and $t^\prime = t\ (\text{or} \ \tau=0)$
correspond to the radiation from the first brane and
the second one, respectively.
When the radiation occurs from the middle points
between the branes, we have $\tau, t^\prime > 0$.

Now we apply the Wick's rotation
$\tau \to -i\tau$.
Using the explicit forms of the
oscillating part and ghost part of
the boundary state, i.e.
Eqs. \eqref{2.4} and \eqref{2.6},
we obtain the following results
\begin{eqnarray}
\label{3.9}
Z^{\rm (osc,g)} &=& \prod_{n=1}^{\infty}
\det \left[ Q^{\dagger}_{(n)  1} Q_{(n)  2}\right]^{-1} \;
\prod_{n=1}^{\infty} \dfrac{(1-q^{2n})^2}{\det
\left({\mathbf 1}- \mathcal{S}^{(1)\dagger}_{(n)}
\mathcal{S}^{(2)}_{(n)} q^{2n} \right) }\;,
\end{eqnarray}
\begin{eqnarray}
\langle e^{ip\cdot X_{\rm osc}} \rangle
&=& \exp \Biggl\{ \frac{\alpha^\prime}{2}
p_\mu p_\nu \sum_{n=1}^{\infty}\frac{1}{n}\Bigg(
\mathcal{S}^{(1)
\dagger\mu\eta}_{(n)}\mathcal{S}^{(2)\nu}_{(n) \ \eta}
\text{Tr}\left[ \ln  \left( \mathbf{1}
- \mathcal{S}^{(1) \dagger}_{(n)}
\mathcal{S}^{(2)}_{(n)}\; q^{2n}
\right)\right]
\nonumber\\
&&\qquad\qquad\qquad - \mathcal{S}^{(2) \mu\nu}_{(n)}
\text{\textbf{T}}_{t'}^{(n)}
- \mathcal{S}^{(1)\dagger \mu\nu}_{(n)}
\text{\textbf{T}}_\tau^{(n)} \Bigg) \Biggr\},\label{3.10}
\end{eqnarray}
where $q=e^{-2(\tau + t')}$, and
\begin{eqnarray}
\text{\textbf{T}}_{(t', \tau)}^{(n)}
= \text{Tr} \left[ \ln \left( \mathbf{1}
- \mathcal{S}^{(1) \dagger}_{(n)}\mathcal{S}^{(2)}_{(n)}
q^{2(n-1)} e^{-4(t^\prime, \tau)}\right)
- \left( \mathbf{1}- \mathcal{S}^{(1) \dagger}_{(n)}
\mathcal{S}^{(2)}_{(n)}
q^{2(n-1)} e^{-4(t^\prime, \tau)}\right)^{-1}\right].
\label{3.11}
\end{eqnarray}
The exponential correlator was calculated
by using the Cumulant expansion in the
eikonal approximation.
For the one-derivative correlators we obtain
the following expression
\begin{eqnarray}
\label{3.12}
\langle \bar{\partial}X^\mu X^\nu\rangle_{\rm osc}
= -\langle \partial X^\mu X^\nu \rangle_{\rm osc}
&=&- i \alpha^\prime \sum_{n=1}^{\infty}
\Bigg\{\eta^{\mu\nu}\text{Tr} \left(
\dfrac{{\mathbf 1}
+ \mathcal{S}^{(1) \dagger}_{(n)}\mathcal{S}^{(2)}_{(n)}q^{2n}}
{{\mathbf 1} - \mathcal{S}^{(1) \dagger}_{(n)}
\mathcal{S}^{(2)}_{(n)} q^{2n}}\right)
\nonumber \\
&-& \mathcal{S}^{(1) \dagger\mu\eta}_{(n)}
\mathcal{S}^{(2)\nu}_{(n) \ \eta}
\text{Tr} \left( \dfrac{ \mathcal{S}^{(1) \dagger}_{(n)}
\mathcal{S}^{(2)}_{(n)}
q^{2n}}{{\mathbf 1} - \mathcal{S}^{(1) \dagger}_{(n)}
\mathcal{S}^{(2)}_{(n)} q^{2n}}\right)
\nonumber\\
&-& \dfrac{1}{4} \left( \mathcal{S}^{(2) \mu\nu}_{(n)}
\partial_{t'} \text{\textbf{T}}_{t'}^{(n)}
-\mathcal{S}^{(1) \dagger \mu\nu}_{(n)}
\partial_{\tau} \text{\textbf{T}}_\tau^{(n)}\right)
\Bigg\}. \ \ \ \ \ \ \ \
\end{eqnarray}
According to the equation
$\langle X^\mu \partial \bar{\partial} X^\nu \rangle = 0$,
the correlator
$\langle \partial X^\mu
\bar{\partial} X^\nu \rangle_{\rm osc}$
in Eq. \eqref{3.3} can be obtained
by computing the derivative of Eq. \eqref{3.12}.

In fact, the integrand of the amplitude
\eqref{3.3} is independent of the worldsheet coordinate
$\sigma$ (see Eqs. \eqref{3.4}-\eqref{3.6} and
\eqref{3.9}-\eqref{3.12}). Therefore,
the integration over the coordinate $\sigma$ is trivial,
i.e. its effect is given by multiplying the amplitude 
by the factor $\pi$.

\subsection{Radiation from the branes with large distance}

We are interested in the radiation of the
closed strings from the branes
whose positions are far from each other.
In the large distance limit (LDL),
the main contribution to the
interaction obviously comes from the exchange
of the massless closed strings.
Since the large distance of the branes
corresponds to the long enough time,
the limit $t \to \infty$
should be exerted on the oscillating
portion of the amplitude,
i.e. on Eqs. \eqref{3.4}-\eqref{3.12}.
Therefore, for the distant branes, the quantity
$Z^{\rm (osc,g)}$ and the
correlators in the amplitude \eqref{3.3}
take the following features
\begin{eqnarray}
Z^{\rm (osc,g)}|_{\rm LDL}
= \prod_{n=1}^{\infty}
\det \left[ Q^{\dagger}_{(n)1}
Q_{(n)2}\right]^{-1},
\label{3.13}
\end{eqnarray}
\begin{eqnarray}
\langle e^{ip\cdot X_{\rm osc}}
\rangle|_{\rm LDL}
&=& \exp \Biggl\{ -\frac{\alpha^\prime}{2}
p_\mu p_\nu \;  \Bigg( \mathcal{S}^{(2) \mu\nu}_{(1)}
\text{\textbf{T}}_{t'}^{(1)}
+ \mathcal{S}^{(1)\dagger\mu\nu}_{(1)}
\text{\textbf{T}}_\tau^{(1)} \Bigg) \Biggr\},
\label{3.14}
\end{eqnarray}
\begin{eqnarray}
\langle \partial X^\mu X^\nu
\rangle_{\rm osc}|_{\rm LDL}
&=& -i \alpha^\prime
\left\lbrace 13 \eta^{\mu\nu}
+ \dfrac{1}{4} \left( \mathcal{S}^{(2)
\mu\nu}_{(1)} \partial_{t'}
\text{\textbf{T}}_{t'}^{(1)}
+ \mathcal{S}^{(1) \dagger \mu\nu}_{(1)} \partial_\tau
\text{\textbf{T}}_{\tau}^{(1)}\right) \right\rbrace
\label{3.15}.
\ \ \ \ \ \ \ \
\end{eqnarray}
Besides, the derivative of Eq. \eqref{3.15} with respect
to $\bar{z}$ may be calculated to obtain the two-derivative
correlator
$\langle \partial X^\mu \bar{\partial} X^\nu
\rangle_{\rm osc}|_{\rm LDL}$.

Note that, for acquiring Eq. \eqref{3.14} we assumed that the 
momentum of the radiated closed string and the setup parameters satisfy the relation
\begin{eqnarray}
p_\mu p_\nu \sum^\infty_{n=2} \left(S^{(1)*}_{(n)}+S^{(2)}_{(n)}
\right)^{\mu\nu}=0.
\end{eqnarray}
This condition clarifies that for the given setup parameters, the components of the momentum of
the radiated closed string are not independent, i.e. two of them are
specified in terms of the others.

Substitute the foregoing correlators into Eq. \eqref{3.3},
the amplitude for the massless string
radiation from the interaction of the branes
in the large distance limit will be obtained.
Since this form of the amplitude is very long we
do not explicitly write it.

\subsection{Radiation in the presence
of the condition
$\Delta^{(1)\dagger}_{(1)}\Delta^{(2)}_{(1)} =\mathbf{1}$}

From now on, for simplification,
we impose the following condition on the setup parameters
$\Delta^{(1)\dagger}_{(1)}\Delta^{(2)}_{(1)} =\mathbf{1}$.
Thus, for the large distance limit, the two-derivative
correlator can be written in the form
\begin{eqnarray}
\langle \partial X^\mu \bar{\partial}
X^\nu \rangle_{\rm osc}|_{\rm LDL}
= - \langle \partial X^\mu X^\nu
\rangle_{\rm osc}|_{\rm LDL}
\left[ \langle p\cdot \partial X p\cdot X
\rangle_{\rm osc}|_{\rm LDL} +
\dfrac{i \alpha^\prime}{2} (k^2 - k^{\prime 2})\right].
\end{eqnarray}
Substitute this correlator into Eq. \eqref{3.3},
the amplitude for the closed string radiation
from the large distance branes takes the feature
\begin{eqnarray}
\mathcal{A}^{(0)}|_{\rm LDL}
&=&- \frac{\pi}{16}T_p^2 \alpha^{\prime 2}
(2\pi)^{26}
\int_{0}^{\infty}d\tau \int_{0}^{\infty}  d t' \
\mathcal{D}(y_1, y_2) e^{4(t' +\tau)}
\int_{-\infty}^{+ \infty}
\int_{-\infty}^{+ \infty}
\prod_{\alpha=0}^{p} dk^\alpha dk^{\prime \alpha}
\nonumber\\
&\times& \mathfrak{D} (k , k^\prime)
e^{- \alpha'(t'k^2+\tau k^{\prime^2})}
\left[\prod_{n=1}^{\infty}
\det \left[ Q^{\dagger}_{(n)1} Q_{(n)2}\right]^{-1} \right]
\langle e^{ip\cdot X_{\rm osc}} \rangle^\prime
\nonumber\\
&\times& \left\lbrace
A (\partial_{t'}\text{\textbf{T}}'^{(1)}_{t'})^2 +
B (\partial_\tau\text{\textbf{T}}'^{(1)}_{\tau})^2 +
C \partial_{t'}\text{\textbf{T}}'^{(1)}_{t'} +
D \partial_\tau\text{\textbf{T}}'^{(1)}_{\tau} +
E \right\rbrace ,
\label{3.18}
\end{eqnarray}
where $\text{\textbf{T}}_{(t', \tau)}'^{(n)}$ and
$\langle e^{ip\cdot X_{\rm osc}} \rangle^\prime$
indicate Eqs. \eqref{3.11} and \eqref{3.14}
in the presence of the equation
$\Delta^{(1)\dagger}_{(1)}\Delta^{(2)}_{(1)} =\mathbf{1}$.
With the help of integration by part
we obtain the following equations
\begin{eqnarray}
\partial_{\tau}\text{\textbf{T}}'^{(1)}_{\tau}
&=& - \dfrac{ 2(\alpha^\prime k^{\prime 2} - 4)}
{\alpha^\prime p_\mu p_\nu
\mathcal{S}^{(1)\dagger \mu\nu}_{(1)}},\\ 
\partial_{t'}\text{\textbf{T}}'^{(1)}_{t'}
&=& - \dfrac{2(\alpha^\prime k^{ 2} -4)}
{ \alpha^\prime p_\mu p_\nu
\mathcal{S}^{(2) \mu\nu}_{(1)}}.
\label{3.19}
\end{eqnarray}

In fact, Eq. \eqref{3.18} formally describes any
massless string radiation. Each of these radiations
possesses its own quantities $A$, $B$, $C$, $D$ and $E$.
In the next section the explicit forms of these
variables for the graviton and dilaton radiations will be
written.

\section{Graviton and dilaton radiations}
\label{400}

In this section, we shall focus on the graviton and dilaton
emission from the interacting distant branes. Precisely,
the amplitude \eqref{3.18} will be rewritten
for the graviton and dilaton radiations, i.e.,
the explicit forms of
the variables $A$, $B$, $C$, $D$ and $E$ will be written.

\subsection{The amplitude for the graviton radiation}

The graviton's polarization tensor is
symmetric and traceless, i.e.,
$\epsilon_{\mu\nu}^{\rm (gr)}
=\epsilon_{\nu\mu}^{\rm (gr)}$ and
$\epsilon_{\quad \ \mu}^{\rm (gr)\mu} =0$.
In addition, the polarization tensor should satisfy
$p^\mu \epsilon_{\mu\nu}^{\rm (gr)} =0  $.
Performing some heavy calculations, for
the graviton radiation we
obtain
\begin{eqnarray}
A &=& \epsilon_{\alpha\beta}^{\rm (gr)}\bigg{\{}
p_\lambda p_\eta \left[ \Delta^{(2) \alpha \lambda}_{(1)}
\Delta^{(2) \beta \eta}_{(1)}
-\Delta^{(2) \alpha \beta}_{(1)}
\Delta^{(2) \lambda \eta}_{(1)}\right]
+ p_{\perp}^2  \Delta^{(2) \alpha \beta}_{(1)}
\nonumber \\
&+& 2 p^\alpha p_\lambda \Delta^{(2) \beta \lambda}_{(1)}
+p^\alpha p^\beta \bigg{\}}
+\epsilon^{\rm{(gr)}\alpha}_{\quad\ \ \alpha}
\left( p^2_\perp + p_\lambda p_\eta
\Delta^{(2) \lambda \eta}_{(1)} \right) ,
\end{eqnarray}
\begin{eqnarray}
B &=& \epsilon_{\alpha\beta}^{\rm (gr)}\bigg{\{}
p_\lambda p_\eta \left[ \Delta^{(1)\dagger
\alpha \lambda}_{(1)}
\Delta^{(1)\dagger \beta \eta}_{(1)}
-\Delta^{(1)\dagger \alpha
\beta}_{(1)}  \Delta^{(1)\dagger
\lambda \eta}_{(1)}\right]
+ p_{\perp}^2  \Delta^{(1)\dagger \alpha \beta}_{(1)}
\nonumber\\
&+& 2 p^\alpha p_\lambda \Delta^{(1)\dagger
\beta \lambda}_{(1)}+p^\alpha p^\beta\bigg{\}}
+\epsilon^{\rm{(gr)}\alpha }_{\quad\ \ \alpha}
\left( p^2_\perp
+ p_\lambda p_\eta \Delta^{(1)\dagger
\lambda \eta}_{(1)} \right),
\end{eqnarray}
\begin{eqnarray}
C &=& \epsilon_{\alpha\beta}^{\rm (gr)}
\bigg{\{} \dfrac{1}{2}(k^2 -k^{\prime 2})
\Delta^{(2) \alpha \beta}_{(1)}
- k^\alpha p_\lambda \Delta^{(2) \beta \lambda}_{(1)}
- k^\beta p_\lambda \Delta^{(2) \alpha \lambda}_{(1)}
\nonumber\\
&+& 2p^\alpha p^\beta\bigg{\}}
-\dfrac{1}{2}
\epsilon^{\rm{(gr)} \alpha }_{\quad\ \ \alpha}
k^2 k^{\prime 2},
\end{eqnarray}
\begin{eqnarray}
D &=& -\epsilon_{\alpha\beta}^{\rm (gr)} \bigg{\{}
\dfrac{1}{2} (k^2 -k^{\prime 2})
\Delta^{(1)\dagger \alpha \beta}_{(1)} - k^\alpha p_\lambda
\Delta^{(1)\dagger \beta \lambda}_{(1)} - k^\beta p_\lambda
\Delta^{(1)\dagger \alpha \lambda}_{(1)}
\nonumber\\
&+& 2 p^\alpha p^\beta\bigg{\}}
+\dfrac{1}{2}
\epsilon^{\rm{(gr)} \alpha }_{\quad\ \ \alpha}
k^2 k^{\prime 2},
\end{eqnarray}
\begin{eqnarray}
E=\epsilon_{\alpha\beta}^{\rm (gr)} k^\alpha k^\beta.
\end{eqnarray}
Note that all indices
represent the worldvolume directions, i.e.,
they belong to the set $\{0,1,\cdots,p\}$.

\subsection{The amplitude for the dilaton radiation}

In contrast to the graviton's and
Kalb-Ramond's polarization tensors,
we have the explicit form of the dilaton polarization tensor.
In the 26-dimensional spacetime, it
possesses the following form
\begin{eqnarray}
\epsilon_{\mu\nu}^{\phi} = \dfrac{1}{\sqrt{24}}
\left( \eta_{\mu\nu} - p_\mu l_\nu - p_\nu l_\mu\right)
, \qquad p\cdot l =1, \qquad l^2=0.
\end{eqnarray}
Thus, the radiation amplitude for the dilaton is given by
Eq. \eqref{3.18}, accompanied by the following variables
\begin{eqnarray}
A &=& \dfrac{1}{\sqrt{24}} p_\xi p_\gamma
\Big\lbrace \mathcal{S}^{(2) \mu \xi}_{(1)}
\mathcal{S}^{(2) \ \gamma}_{(1)\mu}
-\mathcal{S}^{(2) \mu}_{(1) \ \mu}
\mathcal{S}^{(2) \xi \gamma}_{(1)}
\nonumber\\
&-& \left( p_\mu l_\nu + l_\mu p_\nu\right)
\left( \mathcal{S}^{(2) \mu \xi}_{(1)}
\mathcal{S}^{(2) \nu \gamma}_{(1)}
- \mathcal{S}^{(2) \mu \nu}_{(1)}
\mathcal{S}^{(2) \xi \gamma}_{(1)}\right) \Big\rbrace ,\\
B &=& \dfrac{1}{\sqrt{24}} p_\xi p_\gamma
\Big\lbrace \mathcal{S}^{(1)\dagger \mu \xi}_{(1)}
\mathcal{S}^{(1)\dagger \ \gamma}_{(1) \mu}
-\mathcal{S}^{(1)\dagger \mu}_{(1) \;\;\;\mu }
\mathcal{S}^{(1)\dagger \xi \gamma}_{(1)}
\nonumber\\
&-& \left( p_\mu l_\nu
+ l_\mu p_\nu\right)
\left( \mathcal{S}^{(1)\dagger \mu \xi}_{(1)}
\mathcal{S}^{(1)\dagger \nu \gamma}_{(1)}
- \mathcal{S}^{(1)\dagger \mu \nu}_{(1)}
\mathcal{S}^{(1)\dagger \xi \gamma}_{(1)}
\right)\Big\rbrace, \\
C&=& -\dfrac{1}{\sqrt{24}}\bigg\lbrace 2 k^\mu p_\nu
\mathcal{S}^{(2) \ \nu}_{(1)\mu }
- \dfrac{1}{2} (k^2 - k^{\prime 2})
\mathcal{S}^{(2) \mu}_{(1)\ \mu }
- \mathcal{S}^{(2) \mu \nu}_{(1)}
\Big[ p_\mu p_\nu \left( 195 + 2 l\cdot k\right)
\nonumber\\
&+& 2 l_\mu p_\nu \;k\cdot p  - \dfrac{1}{2}
\left( l_\mu p_\nu + p_\mu l_\nu \right)
(k^2 - k^{\prime 2}) \Big]\bigg\rbrace,\\
D&=& \dfrac{1}{\sqrt{24}}\bigg\lbrace 2 k^\mu p_\nu
\mathcal{S}^{(1)\dagger \ \nu}_{(1)\mu}
- \dfrac{1}{2} (k^2 - k^{\prime 2})
\mathcal{S}^{(1)\dagger \mu}_{(1)\;\;\; \mu}
- \mathcal{S}^{(1)\dagger \mu \nu}_{(1)}
\Big[ p_\mu p_\nu \left( 195 + 2 l\cdot k\right)
\nonumber\\
&+& 2 l_\mu p_\nu \;k\cdot p  - \dfrac{1}{2}
\left( l_\mu p_\nu
+ p_\mu l_\nu \right)  (k^2 - k^{\prime 2})
\Big]\bigg\rbrace,\\
E &=& \dfrac{1}{\sqrt{24}}\left[ k^2
- 2 \ (l\cdot k)(p\cdot k) - 28\  k\cdot p - 111\
(k^2- k^{\prime 2})\right].
\end{eqnarray}
Note that all indices in these parameters
are the spacetime indices. They belong
to the set $\{0,1,\cdots, 25\}$.

\subsection{Some physical properties}

Here, by performing the integration over the
proper times, we rewrite the radiation
amplitude \eqref{3.18}. Assume that the momentum
of the radiated string is small. Afterward,
by expanding the exponential, we obtain
\begin{eqnarray}
\int_{0}^{\infty}d\tau \int_{0}^{\infty}
d t' e^{-t' (\alpha^\prime k^2 - 4)}
e^{- \tau (\alpha^\prime k^{\prime^2} - 4)}
\langle e^{ip\cdot X_{\rm osc}} \rangle^\prime
\approx \dfrac{1}{(\alpha^{\prime} k^2-4)
(\alpha^\prime k^{\prime 2}-4)}.
\end{eqnarray}

Now apply a constant shift to the momenta, i.e.,
$k^\alpha=\mathcal{K}^\alpha - u^{\alpha}$ and
$k^{\prime\alpha}= \mathcal{K}^{\prime\alpha}
-u^{\alpha}$. Besides, let the worldvolume vector
$u^\alpha$ satisfy the following conditions
\begin{eqnarray}
k\cdot u = k^\prime\cdot u=0,\;\;\;
u^2= -\frac{4}{\alpha^\prime}.
\end{eqnarray}
Adding all these together, we obtain the final version
of the amplitude as in the following
\begin{eqnarray}
\label{4.14}
\mathcal{A}^{(0)}|_{\rm LDL}
&=& \dfrac{\pi}{4} (2\pi)^{26}T_p^2
\mathcal{D}(y_1, y_2)
\prod_{n=1}^{\infty}\left[
\det\left( Q^{\dagger}_{(n)1} Q_{(n)2}\right)
\right]^{-1}
\nonumber\\
&\times& \int_{-\infty}^{+ \infty}
\int_{-\infty}^{+ \infty}
\prod_{\gamma=0}^{p} d\mathcal{K}^\gamma
d\mathcal{K}^{\prime \gamma}
\mathfrak{D} (\mathcal{K},\mathcal{K}^\prime)
\left(\dfrac{\Gamma}{\mathcal{K}^{2}}
- \dfrac{\Theta}{\mathcal{K}^2 \mathcal{K}^{\prime 2}}
+ \dfrac{\Upsilon}{\mathcal{K}^{\prime 2}} \right),
\end{eqnarray}
where $\Gamma$, $\Upsilon $ and $\Theta$ are given by
\begin{eqnarray}
&~&\Gamma = \dfrac{1}{2 \alpha^\prime p_\mu p_\nu
\mathcal{S}^{(2) \mu\nu}_{(1)}}
\left( D^\prime - \dfrac{B
\mathcal{K}^{\prime 2}}{2 \alpha^\prime p_\mu p_\nu
\mathcal{S}^{(2) \mu\nu}_{(1)}}\right),
\nonumber\\
&~&\Upsilon = -\dfrac{1}{2 \alpha^\prime p_\mu p_\nu
\mathcal{S}^{(1)\dagger \mu\nu}_{(1)}}
\left( C^\prime -
\dfrac{A \mathcal{K}^2}{2 \alpha^\prime p_\mu p_\nu
\mathcal{S}^{(1)\dagger \mu\nu}_{(1)}}\right),
\nonumber\\
&~&\Theta = E^\prime,
\end{eqnarray}
where $ (D^\prime,C^\prime,E^\prime)
\equiv (D,C,E)|_{k \rightarrow
\mathcal{K} - u}^{k^\prime \rightarrow
\mathcal{K}^\prime - u}$.
The amplitude \eqref{4.14} elaborates
the graviton or dilaton
radiation from the interaction of two parallel
dressed-dynamical unstable D$p$-branes
in the large distance. Note that we applied
only one vertex operator, hence,
we do not have simultaneous
radiation of the graviton and dilaton.
However, the three terms in this amplitude clarify
that the graviton and or
dilaton can be radiated in the three
distinct physical processes.
The squared momenta in the denominators, i.e.
$\mathcal{K}^2 $ and $\mathcal{K}^{\prime 2}$,
correspond to the propagators of the
radiated strings by the D-branes. The quantities
$\Gamma$ and $\Upsilon$ correspond to the residue of a
single-pole process. That is, a massless state
is emitted by one of the branes and is absorbed
by the other one, then
after traveling as an excited state,
it re-decays by emitting the final massless string.
The $\Theta$-term exhibits a double-pole process,
in which the radiation occurs between the
branes, from the mediated closed string.
The mediated string, which is exchanged
between the branes, is responsible for the
interaction of the branes.

\section{Conclusions}
\label{500}

We calculated a general amplitude for radiating
any massless closed string via the interaction
of two parallel unstable D$p$-branes. The branes have
tangential dynamics, and they have been dressed by
the Kalb-Ramond field, a quadratic tachyonic field
and a $U(1)$ gauge potential in a particular gauge.
The presence of the background
fields and dynamical parameters
drastically dedicated a generalized form to the amplitude.
These parameters enable us to adjust the value of
the radiation amplitude to any desirable value.

We deformed the foregoing general amplitude to extract
any massless string radiation from the
branes with the large distance. Then, the amplitude for
the graviton and dilaton radiations from the
branes with the large distance
were explicitly computed.
By evaluating the proper-time integrals
in the eikonal approximation, it was revealed that
three distinct radiation processes can occur. That is,
the gravitational or dilatonic radiation can
take place from the first brane, from the second brane
and from the middle region between the branes.


\end{document}